\documentclass[sn-vancouver,Numbered,pdflatex]{sn-jnl}

\usepackage{graphicx}%
\usepackage{multirow}%
\usepackage{amsmath,amssymb,amsfonts}%
\usepackage{amsthm}%
\usepackage{mathrsfs}%
\usepackage[title]{appendix}%
\usepackage{xcolor}%
\usepackage{textcomp}%
\usepackage{manyfoot}%
\usepackage{booktabs}%
\usepackage{algorithm}%
\usepackage{algorithmicx}%
\usepackage{algpseudocode}%
\usepackage{listings}%

\theoremstyle{remark}
\theoremstyle{definition}

\usepackage{float}
\let\origfigure\figure
\let\endorigfigure\endfigure
\renewenvironment{figure}[1][2] {
    \expandafter\origfigure\expandafter[H]
} {
    \endorigfigure
}
\usepackage{lscape}
\usepackage{pdflscape}
\newcommand{\blandscape}{\begin{landscape}}
\newcommand{\elandscape}{\end{landscape}}
\usepackage[none]{hyphenat}
\usepackage[export]{adjustbox}
\usepackage{multirow}
\usepackage{longtable}
\usepackage{array}
\usepackage{wrapfig}
\usepackage{colortbl}
\usepackage{booktabs} 
\usepackage{colortbl}
\usepackage{pdflscape}
\usepackage{tabu}
\usepackage{threeparttable}
\usepackage[normalem]{ulem}
\usepackage{threeparttablex}
\usepackage{makecell}

\usepackage{booktabs}
\usepackage{longtable}
\usepackage{array}
\usepackage{multirow}
\usepackage{wrapfig}
\usepackage{float}
\usepackage{colortbl}
\usepackage{pdflscape}
\usepackage{tabu}
\usepackage{threeparttable}
\usepackage{threeparttablex}
\usepackage[normalem]{ulem}
\usepackage{makecell}
\usepackage{xcolor}

\usepackage{longtable,booktabs,array}
\usepackage{calc} 
\usepackage{etoolbox}
\makeatletter
\patchcmd\longtable{\par}{\if@noskipsec\mbox{}\fi\par}{}{}
\makeatother
\IfFileExists{footnotehyper.sty}{\usepackage{footnotehyper}}{\usepackage{footnote}}
\makesavenoteenv{longtable}

\begin{document}

\title[Framework for transferable computational models]{A prototype software framework for transferable computational health economic models and its early application in youth mental health}


\author*[1,2]{\fnm{Matthew} \spfx{P} \sur{Hamilton} }\email{\href{mailto:matthew.hamilton1@monash.edu}{\nolinkurl{matthew.hamilton1@monash.edu}}}

\author[2,3,1]{\fnm{Caroline} \sur{Gao} }

\author[4]{\fnm{Glen} \sur{Wiesner} }

\author[2,3]{\fnm{Kate} \spfx{M} \sur{Filia} }

\author[2,3]{\fnm{Jana} \spfx{M} \sur{Menssink} }

\author[5]{\fnm{Petra} \sur{Plencnerova} }

\author[2,3]{\fnm{David} \spfx{G} \sur{Baker} }

\author[2,3]{\fnm{Patrick} \spfx{D} \sur{McGorry} }

\author[6,3]{\fnm{Alexandra} \sur{Parker} }

\author[7]{\fnm{Jonathan} \sur{Karnon} }

\author[2,3]{\fnm{Sue} \spfx{M} \sur{Cotton} }

\author[1]{\fnm{Cathrine} \sur{Mihalopoulos} }

  \affil*[1]{\orgdiv{School of Public Health and Preventive Medicine}, \orgname{Monash University}, \orgaddress{\city{Melbourne}, \country{Australia}}}
  \affil[2]{\orgname{Orygen}, \orgaddress{\city{Parkville}, \country{Australia}}}
  \affil[3]{\orgdiv{Centre for Youth Mental Health}, \orgname{The University of Melbourne}, \orgaddress{\city{Parkville}, \country{Australia}}}
  \affil[4]{\orgname{Heart Foundation}, \orgaddress{\city{Docklands}, \country{Australia}}}
  \affil[5]{\orgname{headspace National Youth Mental Health Foundation}, \orgaddress{\city{Melbourne}, \country{Australia}}}
  \affil[6]{\orgdiv{Institute for Health and Sport}, \orgname{Victoria University}, \orgaddress{\city{Footscray}, \country{Australia}}}
  \affil[7]{\orgdiv{Flinders Health and Medical Research Institute}, \orgname{Flinders University}, \orgaddress{\city{Adelaide}, \country{Australia}}}

\abstract{\textbf{Summary: } We are developing an economic model to explore multiple topics in Australian youth mental health policy. We want that model to be readily transferable to other jurisdictions. We developed a software framework for authoring transparent, reusable and updatable Computational Health Economic Models (CHEMs) (the software files that implement health economic models). We specified framework user requirements of a template CHEM module that facilitates modular model implementations, a simple programming syntax and tools for authoring new CHEM modules, supplying CHEMs with data, reporting reproducible CHEM analyses, searching for CHEM modules and maintaining a CHEM project website. We implemented the framework as six development version code libraries in the programming language R that integrate with online services for software development and research data archiving. We used the framework to author five development version R libraries of CHEM modules focused on utility mapping in youth mental health. These modules provide tools for variable validation, dataset description, multi-attribute instrument scoring, construction of mapping models, reporting of mapping studies and making out of sample predictions. We assessed these CHEM module libraries as mostly meeting transparency, reusability and updatability criteria that we have previously developed, but requiring more detailed documentation and unit testing of individual modules. Our software framework has potential value as a prototype for future tools to support the development of transferable CHEMs. \newline \newline \textbf{Code: } Visit \url{https://www.ready4-dev.com} for more information about how to find, install and apply the prototype software framework. \newline \newline}



\maketitle

\textbf{Accepted for publication:}

\begin{itemize}
\item
  This manuscript has been accepted for publication at a scientific journal.
\item
  Before citing, please check article metadata for details about where to find the final manuscript.
\end{itemize}

\hypertarget{introduction}{%
\section{Introduction}\label{introduction}}

Computational models, particularly those addressing economic topics, have become essential tools for health policy development \citep{dakin2015influence, Erdemir2020}. Although influential and widely used, health economic models can be skills and resource intensive to develop. One suggestion for making health economic modelling projects more tractable is for health economists to re-use each other's models \citep{Arnold2010}.

Key concerns for health economists when considering whether to reuse a model are generalisability (application without adaptation) and transferability (application of selected components and/or application with modification) \citep{RN39}. Health economic models are likely to require at least some modification before they can be validly applied to a new jurisdiction, time period or decision problem. Making such modifications is harder if the software files that implement the model - the Computational Health Economic Model (CHEM) -- cannot be readily understood, used and edited.

Some of the barriers to the development of transferable CHEMs could be addressed by appropriate software frameworks. A software framework is a shared common technology used by developers to collaboratively author software \citep{myllarniemi2018development}. A software framework provides a foundation for developing multiple software applications with shared resources (e.g., code and data files), that can be modified to suit specific needs. Advantages of using software frameworks include facilitating code reuse and extension, promoting good programming practice and the capability to provide enhanced functionality and performance without additional effort by developers \citep{edwin2014software}.

In this paper, we describe:

\begin{enumerate}
\def\labelenumi{(\roman{enumi})}
\item
  our motivation for developing a transferable CHEM in youth mental health;
\item
  a prototype software framework we have developed to support the development of transferable CHEMs; and
\item
  how we have applied the prototype software framework to undertake initial development of our youth mental health CHEM.
\end{enumerate}

\hypertarget{motivation}{%
\section{Motivation}\label{motivation}}

We became interested in solving some of the technical barriers to transferable CHEMs in 2018 when considering how to synthesise multiple youth mental health economic modelling projects in one multi-purpose model. This overarching youth mental health model is called readyforwhatsnext \citep{rfwn2024}.

\hypertarget{economic-topics}{%
\subsection{Economic topics}\label{economic-topics}}

We are constructing readyforwhatsnext from projects in four of the twelve domains of health economics identified by Wagstaff and Culyer \citep{wagstaff2012four}. These are:

\begin{itemize}
\item
  health and its value (our projects: models to map psychological and functional measures to health utility);
\item
  determinants of health and ill-health (our projects: models for creating synthetic household populations with risk and protective factors for mental disorders);
\item
  demand for health and health care (our projects: spatial epidemiology and help-seeking choice models); and
\item
  supply of health services (our projects: a model of primary mental health care services for young people).
\end{itemize}

Potential future directions for readyforwhatsnext may incorporate projects in two additional domains: (i) public health (to model the impact of selected fiscal policy and regulation options on young people's mental health); and (ii) human resources (to model the supply and behaviours of the youth mental health workforce). Our ultimate aim is to flexibly combine model components in analyses that answer questions in a further two domains:

\begin{itemize}
\item
  efficiency and equity (our goal: assess the distributional impacts and identify the optimal targeting of youth mental health interventions); and
\item
  economic evaluation (our goal: assess the value for money and affordability of competing policy options for improving the mental health of young people).
\end{itemize}

\hypertarget{transferability}{%
\subsection{Transferability}\label{transferability}}

We are principally interested in using readyforwhatsnext to answer policy questions relating to the mental health of young people in Australia. Australian public mental health services can be planned and commissioned by the Australian Government, the governments of Australia's state and territories and independent, regionally focused organisations called Primary Health Networks. The mental health of young people is a global priority \citep{https://doi.org/10.1002/wps.20938} and Australia has initiated youth mental health service model reforms that have been adopted in other jurisdictions \citep{RN27}. For these reasons, we aim to develop readyforwhatsnext in a manner that facilitates its transfer to multiple decision contexts.

However, there are ethical risks associated with each potential transfer, in particular the need to ensure the social acceptability, fitness for purpose and appropriate use of a CHEM in each new decision context \citep{ethicsCHEMS2024}. We therefore aim for readyforwhatsnext to meet six \emph{transparency}, \emph{reusability} and \emph{updatability} (TRU) criteria. We have described these criteria and their origins in ethical modelling practice elsewhere \citep{ethicsCHEMS2024}, but briefly they are:

T1) Software files are open access;

T2) Developer contributions and judgments on appropriate use are easily identified;

R1) Programming practices promote selective reuse;

R2) Licenses permit derivative works;

U1) Maintenance infrastructure is in place; and

U6) Releases are systematically retested and deprecated.

To meet these criteria we are implementing readyforwhatsnext as an open-source, modular model. Open source projects grant third parties permissions to access, use and modify model source code and data. A modular model is constructed from multiple reusable and replaceable sub-models (modules) \citep{pan2021modular}. When undertaking analyses intended to inform decision-making, readyforwhatsnext uses real Australian data (which can be empirical, simulated or assumption-based). To help demonstrate the potential transferability of readyforwhatsnext to other decision contexts, we are creating synthetic or toy datasets (``fake'' data that realistically represent the key features of observed data, but which can be shared without confidentiality concerns \citep{bellovin2019privacy}).

A high-level schematic guiding our implementation of readyforwhatsnext is outlined in Figure \ref{fig:fig1}. To implement this plan, we first developed a software framework to assist with general (i.e., model agnostic) tasks relating to authoring, management and use of CHEM modules, datasets and analysis and reporting programs. We then used that framework to develop artefacts (e.g., variable validation tools) specific to the distinctive features of the readyforwhatsnext model.

\bgroup 
    \origfigure[H]

\includegraphics[width=400px,]{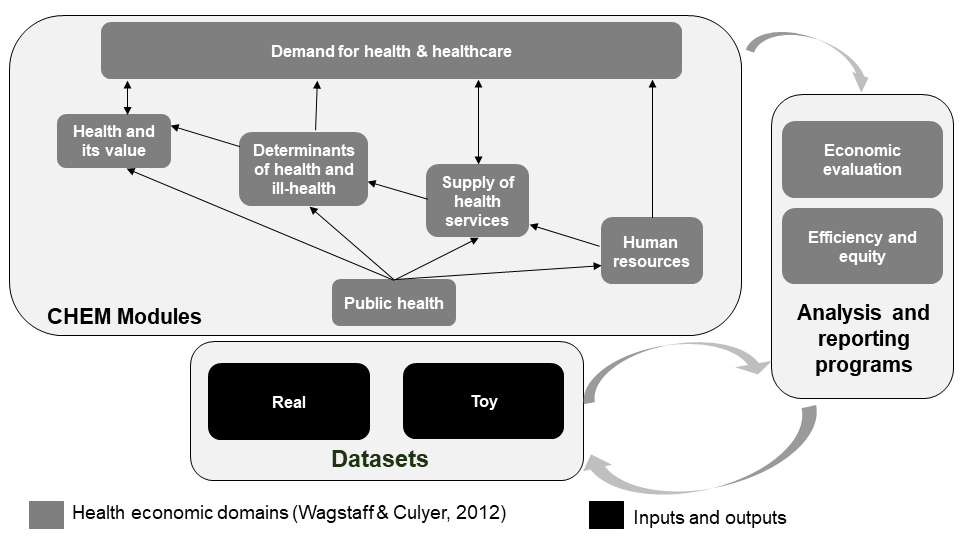} \caption{High level summary of planned implementation of youth mental health economic model}\label{fig:fig1}

    \endfigure
\egroup

\hypertarget{prototype-software-framework}{%
\section{Prototype software framework}\label{prototype-software-framework}}

To help readyforwhatsnext meet all six TRU criteria, we developed a prototype software framework called ready4 \citep{ready424}.

\hypertarget{framework-user-requirements}{%
\subsection{Framework user requirements}\label{framework-user-requirements}}

The framework user requirements (FURs) for ready4 evolved over six years (2018-2024) as we encountered new implementation issues for readyforwhatsnext that we needed to solve. Currently, these requirements are:

FUR 1) A reusable \emph{template} for CHEM modules that can be: (i) run as independent models; (ii) safely combined, sharing inputs and outputs with each other without unintended interference; and (iii) selectively replaced, updated or extended.

FUR 2) A \emph{programming syntax} that promotes simplicity and consistency in how algorithms associated with CHEM modules are: (i) labelled; and (ii) used in health economic analysis programs.

FUR 3) \emph{Module authoring tools} that integrate with online software development services to help author CHEM modules that are: (i) created using the template module; (ii) written in a consistent house style; (iii) always at least minimally documented; (iv) licensed for reuse; (v) bundled as R libraries with authorship, funding source, version number and Digital Object Identifier (DOI) metadata; and (vi) quality assured with publicly available testing artefacts.

FUR 4) \emph{Data management tools} that integrate with online data repository services to streamline: (i) the ingesting and labeling of confidential and non-confidential data; and (ii) the sharing of non-confidential data.

FUR 5) \emph{Replication and reporting tools} that: help (i) author analysis and reporting template programs for rendering HTML, Adobe PDF and Microsoft Word outputs; (ii) retrieve version specific analysis and reporting template programs from online code repositories and (iii) supply analysis and reporting template programs with study specific data (e.g., authorship details, input data and analysis results) to render custom reports.

FUR 6) \emph{Search tools} to retrieve web-based information on a CHEM project's: (i) model modules and tutorials; (ii) datasets and data collections; and (iii) replication and reporting template programs.

FUR 7) \emph{Documentation tools} that integrate with open source website development and hosting infrastructure to help: (i) develop a CHEM project documentation website that consolidates content from the websites of individual CHEM module libraries, datasets and programs; and (ii) partially automate the updating of website resources such as tutorials and itemised lists of code and data releases.

\hypertarget{framework-implementation}{%
\subsection{Framework implementation}\label{framework-implementation}}

We implemented ready4 as R \citep{RCORE2022} code libraries that integrate with a number of online services. The framework is documented by a project website.

\hypertarget{r-libraries}{%
\subsubsection{R libraries}\label{r-libraries}}

R libraries typically depend on other R libraries. As the number of dependencies of an R library grows, so does the fragility of that library (e.g., the library may cease to work as intended due to changes in one of its dependency libraries). To reduce the fragility of our framework we implemented it as multiple R libraries rather than one R library. We authored six novel R libraries to implement the ready4 framework, all of which have distinct purposes and dependencies (Table \ref{tab:cpkgs}).

To meet \emph{FUR 1}, the ready4 \citep{ready4lib24} foundation library defines a template CHEM module using R's S4 class system. Benefits of using the S4 class system include automated verification that data supplied to modules adhere to pre-specified criteria and support for object-oriented programming. Two notable features of object-oriented programming are encapsulation \citep{8717448} and inheritance \citep{8717448}. Encapsulation allows a CHEM module to bundle model data and algorithms in a protected environment, which is useful for helping ensure that CHEM modules continue to work as intended when combined. Inheritance allows new CHEM modules to inherit some or all of the properties of a parent module. Inheritance is useful when modifying CHEM modules, for example to account for: (i) interaction effects when CHEMs are combined; or (ii) distinctive features of a new decision contexts to which they are being transferred. The ready4 template CHEM module can be used to create other CHEM modules that inherit a common set of properties. One of these inherited properties is a novel syntax of 15 core commands \citep{ready4syntax24} that enable CHEM module algorithms to be consistently named and described (meeting \emph{FUR 2}). The ready4 library also contains tools for retrieving web-based information on CHEM modules, datasets and analysis programs (meeting \emph{FUR 6}) and tools for developing and maintaining a CHEM project's documentation website (meeting \emph{FUR 7}).

Three module authoring libraries are designed to meet \emph{FUR 3}. The ready4fun \citep{ready4funlib24} library contains tools for writing functions in a consistent house style and automatically generating basic documentation for each function. The ready4class \citep{ready4classlib24} library contains tools to help streamline and standardise the authoring of novel, documented CHEM modules from the template CHEM module. The ready4pack \citep{ready4packlib24} library provides tools for disseminating themed bundles of CHEM modules as R libraries that are: documented (with a website and PDF manuals); licensed (using the copyleft GNU GPL-3 \citep{GNUGPL2022} by default); easily citable (citation information can be retrieved within an R session or from hosting repositories); and quality assured (each update triggers the running of any tests created by module library authors, with test results publicly documented).

The ready4use \citep{ready4uselib24} library for managing CHEM data contains tools for supplying CHEM modules with input data ingested from local (i.e., a user's computer) or remote (online repositories) locations, labelling CHEM module datasets and exporting CHEM module data to online repositories (meeting \emph{FUR 4}). The ready4show \citep{ready4showlib24} library for authoring reproducible analysis programs and reporting subroutines (\emph{FUR5}) contains tools to help write executables that apply CHEM modules to compatible datasets for the purpose of undertaking reproducible health economic analyses. The library supports the creation of analysis programs that are self-documenting (code is integrated with plain English explanations of what it does) and subroutines that can be used as templates that trigger the creation of explanatory documents (e.g., a scientific manuscript).

\begin{table}
\centering\centering
\caption{\label{tab:cpkgs}Software framework R libraries}
\centering
\begin{tabular}[t]{l>{\raggedright\arraybackslash}m{20em}>{\raggedright\arraybackslash}m{20em}}
\toprule
Library & Purpose & R library dependencies\\
\midrule
\cellcolor{gray!10}{ready4} & \cellcolor{gray!10}{Provide a template and novel syntax for modular CHEM implementations and tools for finding interoperable CHEM modules, datasets and reproducible analysis programs.} & \cellcolor{gray!10}{dataverse dplyr gh kableExtra lifecycle magrittr methods piggyback purrr rlang rvest stats stringi stringr tibble tidyRSS tidyselect tools utils}\\
\addlinespace
ready4fun & Streamline and standardise the authoring and documenting of functions that support transferable and generalisable model algorithms. & desc devtools dplyr gert Hmisc lifecycle lubridate magrittr methods piggyback pkgdown purrr readxl ready4 ready4show ready4use rlang sinew stats stringi stringr testit tibble tidyr tools usethis utils xfun\\
\addlinespace
\cellcolor{gray!10}{ready4class} & \cellcolor{gray!10}{Streamline and standardise the authoring and documenting of new interoperable CHEM modules.} & \cellcolor{gray!10}{devtools dplyr fs gtools Hmisc knitr lifecycle magrittr methods purrr ready4 ready4fun ready4show rlang stats stringi stringr testit testthat tibble tidyr usethis utils}\\
\addlinespace
ready4pack & Help bundle and disseminate newly created CHEM modules as R libraries that are documented, licensed and quality assured. & dataverse dplyr knitr lifecycle magrittr methods purrr ready4 ready4class ready4fun rlang stringr testthat tibble tidyr utils\\
\addlinespace
\cellcolor{gray!10}{ready4use} & \cellcolor{gray!10}{Help manage the labelling and transfer of data between CHEM modules and local and remote data repositories.} & \cellcolor{gray!10}{data.table dataverse dplyr fs Hmisc knitr lifecycle magrittr methods piggyback purrr readxl ready4 ready4show rlang stats stringi stringr testit testthat tibble tidyr utils}\\
\addlinespace
ready4show & Facilitate the use of CHEM modules in programs to implement integrated and reproducible data ingest, analysis and reporting pipelines. & dataverse DescTools dplyr flextable grDevices here Hmisc kableExtra knitr knitrBootstrap lifecycle magrittr methods officer purrr ready4 rlang rmarkdown stringi stringr testthat tibble tidyr utils xtable\\
\bottomrule
\end{tabular}
\end{table}

\hypertarget{online-services}{%
\subsubsection{Online services}\label{online-services}}

Framework libraries are designed to be used in conjunction with a number of online services. The module authoring libraries use the GitHub \citep{github2007} service's tools for: software development, testing, version control and distribution; hosting of documentation websites; and transparent record keeping of collaborator contributions. Continuous integration \citep{CI2017} tools provided by GitHub are used to trigger systematic retesting of module libraries with every edit. Tests can be unit tests (verifying the correct output when small, isolated sections of code are run independently) and acceptance tests (verifying the correct output from multiple code components are run together to meet core user-requirements \citep{martin2003agile}). The module authoring libraries use the codecov \citep{codecov_2022} service to measure and share information about code coverage \citep{ERICWONG2010188} - the proportion of code that has been explicitly tested. The open science repository service Zenodo \citep{Zenodo2013} is used by the module authoring libraries to provide persistent storage, uniquely identified with a Digital Object Identifier (DOI), of each code release. The ready4use library uses the Dataverse \citep{Dataverse2007} data repository service for persistent, uniquely identifiable and versioned storage of model input and output data. For the management of frequently updated datasets that do not require persistent storage, all framework libraries use GitHub.

\hypertarget{documentation-website}{%
\subsubsection{Documentation website}\label{documentation-website}}

A project documentation website (\url{https://www.ready4-dev.com}) provides guidance to model developers on how to use and contribute improvements to the ready4 software framework. Prior versions of the documentation website are archived and publicly accessible. The documentation website was developed using the Hugo framework \citep{hugo_2023}, Docsy theme \citep{docsy_2023} and Algolia DocSearch \citep{algoliadocsearch_2023} and is hosted using the Netlify \citep{netlify_2023} service. We linked our Netlify account to our GitHub organisation so that the project website automatically updates whenever its (publicly available) source code is edited.

\hypertarget{prototype-model-modules}{%
\section{Prototype model modules}\label{prototype-model-modules}}

Our early use of the ready4 framework has focused on developing readyforwhatsnext modules for constructing, sharing and using utility mapping models in youth mental health.

\hypertarget{model-user-requirements}{%
\subsection{Model user requirements}\label{model-user-requirements}}

The model user requirements (MURs) that we specified for the readyforwhatsnext's initial modules are:

MUR 1) \emph{Variable validation tools} that: (i) define the allowable values and numeric class (integer or double precision) associated with different psychological, functional and health utility measures appropriate for use in youth mental health samples; (ii) provide an informative error message that halts an analysis or reporting program's execution should impermissible values be assigned to each measure.

MUR 2) \emph{Dataset description tools} that: (i) attach structured metadata to a human records dataset that specify unique person identifier, group assignment and data collection round variables; and (ii) generate tabular and graphical summaries of dataset descriptive statistics.

MUR 3) \emph{Multi-attribute instrument scoring tools} that: (i) attach structured metadata about an instrument (e.g.~name, version, country, items, domains, parameter values for scoring algorithms) to a human records dataset; (ii) apply pre-written (for AQol-6D \citep{Richardson2012AQ} and EQ-5D \citep{devlin2017eq} instruments) or user-supplied instrument scoring functions to a human records dataset; and (iii) generate descriptive plots of a dataset's instrument scores for individual items, domains and totals (weighted and unweighted).

MUR 4) \emph{Utility mapping model construction tools} that: (i) attach structured metadata about the target utility instrument and candidate predictor variables and covariates to be used in constructing utility mapping models to a human records dataset; (ii) generate descriptive tables, a correlation matrix and plots about the variables to be assessed for inclusion in models; (iii) attach structured metadata about the candidate utility mapping models to be assessed; (iv) generate tabular and graphical summaries of the performance of each combination of model, predictors and covariates specified for evaluation by either a user or by a default algorithm; (v) strip confidential records from the constructed models so that they are able to be publicly shared.

MUR 5) \emph{Utility mapping study reporting tools} that: (i) generate from a reporting template a PDF catalogue summarising the predictive performance under multiple configurations of utility mapping models selected for public dissemination; (ii) generate from a reporting template (either a pre-written default or a user-supplied customisation) a scientific manuscript summarising a utility mapping study; (iii) share model catalogues and (if desired) scientific summary along with metadata about utility mapping models in an open access repository; and (iv) facilitate the authoring of a consolidated study replication program that implements data ingest, instrument scoring, model construction, report authoring and dissemination of study artefacts.

MUR 6) \emph{Utility prediction tools} that: (i) search open access repositories for utility mapping models developed with readyforwhatsnext; (ii) retrieve relevant metadata, including a link to the model catalogue for models specified by a user; (iii) apply selected utility mapping models to make out of sample predictions using a dataset provided by a user; and (iv) (where there is health utility data at two time points) convert predicted health utility scores to Quality Adjusted Life Years (QALYs).

\hypertarget{model-implementation}{%
\subsection{Model implementation}\label{model-implementation}}

\hypertarget{online-service-configuration}{%
\subsubsection{Online service configuration}\label{online-service-configuration}}

We established and configured accounts with the online services supported by ready4. We created a GitHub organisation (a collection of code repositories) where source code that we author is stored and version controlled at \url{https://github.com/ready4-dev/}. We configured individual repositories in our GitHub organisation to implement continuous integration to assess the compliance of our module libraries with policies specified by the Comprehensive R Archive Network (CRAN) \citep{CRAN2022}. To facilitate the creation and hosting of module library websites, we enabled GitHub Pages in each code repository. We also created a Zenodo community, a collection of permanent, uniquely identified repositories, at \url{https://zenodo.org/communities/ready4/}. We then linked our Zenodo community and GitHub organisation so that every time we specify a version of code in one of our GitHub repositories as a ``release'', a copy of that code is automatically archived on Zenodo with a DOI. Finally, to manage model datasets, we created a dedicated collection (\url{https://dataverse.harvard.edu/dataverse/ready4}) within the Harvard Dataverse installation.

\hypertarget{module-libraries}{%
\subsubsection{Module libraries}\label{module-libraries}}

We used all six ready4 framework libraries to author five development version readyforwhatsnext module libraries (Table \ref{tab:rfwnlibs}). These libraries are youthvars \citep{hamilton_matthew_2022_6084467} which addresses MURs 1 and 2, scorz \citep{hamilton_matthew_2022_6084824} which addresses MUR 3, specific \citep{hamilton_matthew_2022_6116701} which addresses MUR 4, TTU \citep{gao_caroline_2022_6130155} which addresses MUR 5 and youthu \citep{matthew_p_hamilton_2021_5646669} which addresses MUR 6.

\hypertarget{reporting-templates}{%
\subsubsection{Reporting templates}\label{reporting-templates}}

We used the ready4show library to author two reporting templates that are designed to be used in conjunction with these utility mapping modules. The first reporting template \citep{hamilton_matthew_2022_6116385} is used to create a catalogue of utility mapping models. The second reporting template \citep{matthew_p_hamilton_2022_5976988} can be used to generate a manuscript providing a scientific summary of each study implemented with these modules.

\hypertarget{application}{%
\subsubsection{Application}\label{application}}

We have previously described a study \citep{Hamilton2021.07.07.21260129} that used these module libraries and reporting templates to develop and document utility mapping models in a sample of young people presenting to primary mental health services. We used these module libraries to write programs for replicating that study's reporting and analysis algorithm \citep{hamilton_matthew_2022_6129906}, to demonstrate how to apply the utility mapping models constructed in that study to new data \citep{hamilton_matthew_2022_6416330} and to generate a synthetic representation of the study dataset \citep{hamilton_matthew_p_2022_6321821}. We also created an open access study input parameters and results dataset \citep{DVN/DKDIB0_2021} and toy datasets to help demonstrate the transferability of the study algorithm \citep{DVN/HJXYKQ_2021}.

\begin{table}
\centering\centering
\caption{\label{tab:rfwnlibs}Model module libraries for utility mapping}
\centering
\begin{tabular}[t]{l>{\raggedright\arraybackslash}m{20em}>{\raggedright\arraybackslash}m{20em}}
\toprule
Library & Purpose & R library dependencies\\
\midrule
\cellcolor{gray!10}{youthvars} & \cellcolor{gray!10}{Describe and validate youth mental health human record datasets.} & \cellcolor{gray!10}{arsenal assertthat car cowplot dplyr ggplot2 ggpubr gridExtra here Hmisc hutils kableExtra knitr knitrBootstrap lifecycle lubridate magrittr MASS Matrix matrixcalc methods mice psych purrr ready4 ready4show ready4use rlang scales simstudy stats stringi stringr testthat tibble tidyselect utils}\\
\addlinespace
scorz & Score multi-attribute instruments. & dplyr eq5d Hmisc hutils knitr knitrBootstrap lifecycle magrittr methods mice purrr ready4 ready4use rlang simstudy stats stringi stringr testthat tibble tidyselect youthvars\\
\addlinespace
\cellcolor{gray!10}{specific} & \cellcolor{gray!10}{Construct and evaluate utility mapping models.} & \cellcolor{gray!10}{assertthat boot Boruta brms caret cowplot dataverse DescTools dplyr enrichwith ggalt ggfortify ggplot2 gtools here Hmisc hutils kableExtra knitr knitrBootstrap lifecycle magrittr MASS methods parallel psych purrr randomForest ready4 ready4show ready4use rlang scorz stats stringi stringr synthpop testthat tibble tidyr tidyselect utils viridis xfun youthvars}\\
\addlinespace
TTU & Implement reproducible utility mapping sudies. & dplyr ggplot2 Hmisc knitr lifecycle magrittr methods purrr R.utils ready4 ready4show ready4use rlang scorz specific stringr testthat utils\\
\addlinespace
\cellcolor{gray!10}{youthu} & \cellcolor{gray!10}{Apply utility mapping models to out of sample data.} & \cellcolor{gray!10}{assertthat BCEA boot dataverse dplyr lifecycle lubridate magrittr MatchIt methods purrr ready4 ready4use rlang specific stats stringi stringr tibble tidyr tidyselect truncnorm utils youthvars}\\
\bottomrule
\end{tabular}
\end{table}

\hypertarget{documentation-website-1}{%
\subsection{Documentation website}\label{documentation-website-1}}

Using the framework website as a template, we used the search and documentation tools from the ready4 library to create and maintain a project documentation website (\url{https://readyforwhatsnext.org/}). The project website includes details on available model modules, datasets and analysis programs and tutorials on how these artefacts can be used.

\hypertarget{transferability-assessment}{%
\subsection{Transferability assessment}\label{transferability-assessment}}

We subjectively assessed these utility mapping model artefacts against TRU criteria (Table \ref{tab:checktb}). For each criterion, we provided a global assessment of whether it was met using the responses ``yes'', ``no'' or ``partial''. We believe that our utility mapping study module libraries, code and data repositories and documentation have satisfactorily met four of the six criteria (T1, R1, R2 and U1) and have partially met two criteria (T2 and U2). The main shortcomings we identified were that individual modules and functions have yet to: (i) be supported with human authored documentation that includes examples of how they can be used; and (ii) be quality assured through unit testing.

\begin{landscape}

\begin{table}
\centering\centering
\caption{\label{tab:checktb}Assessment of a CHEM implementation in youth mental health using Transparent, Reusable and Updatable (TRU) criteria}
\centering
\begin{tabular}[t]{>{\raggedright\arraybackslash}p{10em}l>{\raggedright\arraybackslash}p{35em}}
\toprule
Criterion & Met? & Detail\\
\midrule
\cellcolor{gray!10}{\textbf{T1 Software files are open access}} & \cellcolor{gray!10}{Yes} & \cellcolor{gray!10}{All source code and testing procedures are available in public GitHub repositories, with each code release persistently available in a Zenodo repository. As the study dataset contains confidential patient records, it was not published. Instead, a synthetic representation of the study dataset is persistently available in a repository in the Harvard Dataverse. Data files to support out of sample application of models are published at the same location.}\\
\textbf{T2 Developer contributions and judgments on appropriate use are easily identified} & Partial & All module libraries, programs, and datasets are distributed with citation information. Public GitHub repositories detail author code contributions over project development history. Model catalogues persistently available on the Harvard Dataverse describe the predictive performance of models under multiple usage regimes. Each code library is documented with worked examples of how to apply CHEM modules. However, documentation of individual functions and methods are algorithm generated and need more customisation, including more examples of use. Analysis and reporting programs are self-documenting. Sub-routines for generating reports are documented with README files.\\
\cellcolor{gray!10}{\textbf{R1 Programming practices promote selective reuse}} & \cellcolor{gray!10}{Yes} & \cellcolor{gray!10}{CHEMs are written using both functional and object-oriented paradigms. Code library documentation websites include hypothetical examples of generalisability (applying study algorithm to estimate mapping models from new data with the same predictor and outcome variables) and transferability (adapting study algorithm to develop mapping models from datasets with different predictor variables and outcomes measured with a different utility instrument).}\\
\textbf{R2 Licenses permit derivative works} & Yes & All code is distributed using GPL-3 licenses. Datasets use an amended version of template terms provided by Harvard Dataverse, allowing reuse of data subject to some ethical restrictions (e.g., use in efforts to reidentify study participants is prohibited).\\
\cellcolor{gray!10}{\textbf{U1 Maintenance infrastructure is in place}} & \cellcolor{gray!10}{Yes} & \cellcolor{gray!10}{All code is version controlled using git and GitHub, with semantic versioning. Each code library has a specified maintainer and guidance for potential code contributors is available on the project website.}\\
\addlinespace
\textbf{U2 Releases are systematically retested and deprecated} & Partial & Continuous integration is used for all code libraries, primarily for acceptance testing. Only limited use is made of unit testing. Retired library code is deprecated using tools from the lifecycle R library. Library documentation articles and datasets are also deprecated.\\
\bottomrule
\end{tabular}
\end{table}

\end{landscape}

\hypertarget{discussion}{%
\section{Discussion}\label{discussion}}

We have developed a software framework and used it to author CHEM modules that largely satisfy explicit criteria for transparency, reusability and updatability. We have used toy data to demonstrate how these CHEM module libraries, originally developed for a single utility mapping study \citep{Hamilton2021.07.07.21260129}, can be applied: (i) to datasets from new samples with different concepts and variable names; (ii) to construct models with different utility instruments, predictors and covariates; and (iii) to develop and use both cross-sectional and longitudinal mapping models.

Our framework and CHEM module R libraries are currently publicly available as development releases. We plan to undertake a number of key improvements, such as more detailed documentation, comprehensive unit testing and streamlining dependencies, prior to submitting these libraries for quality assurance and dissemination by CRAN \citep{CRAN2022}.

Although designed to address challenges specific to our project, ready4 can be a prototype for future software frameworks that meet the user requirements of other modelling teams and projects. Our prototype has a number of features that such future work can build on. Firstly, developing a software framework to work within an existing and widely used open source programming language such as R, can keep framework scope relatively narrow. This makes it more tractable to develop, maintain and learn, and facilitates integration with other modelling tools written in that language (e.g., the dependency libraries we list in Tables \ref{tab:cpkgs} and \ref{tab:rfwnlibs}). Secondly, a sensible trade-off needs to be found between transparent code implementation (which requires clear and sufficiently detailed documentation) and prioritising the development of working code over writing documentation (a foundation principle of Agile Software Development \citep{beck2001manifesto}). Our software framework makes this trade off by enforcing the use of consistent code style conventions and file organisation which in turn enables automated generation of simple documentation at every CHEM module update.

Our framework addresses barriers to transferability specific to how a model is implemented as software (ie, CHEMs). Health economists must also consider conceptual or theoretical validity. For example, the candidate models currently available in our scorz library \citep{hamilton_matthew_2022_6084824} were chosen because of their suitability for mapping to the utility instrument (AQoL-6D \citep{Richardson2012AQ}) used in the study \citep{Hamilton2021.07.07.21260129} for which scorz was originally developed. We have identified a need to use alternative models for mapping to utility instruments that produce less continuous total score distributions \citep{Hamilton2021.07.07.21260129}. Our software framework is designed to make such extensions more straightforward to implement.

Factors such as usability, active user-community and supporting resources influence adoption of software frameworks \citep{myllarniemi2018development}. To systematically address these issues for our prototype would require investments in: (i) engaging the health economic modelling community in refining framework user requirements; (ii) undertaking software re-development and testing to meet those requirements: (iii) incoporating the evolving code and documentation authoring capabilities of large language models; (iv) the development of improved and more comprehensive documentation and training resources; and (v) proactive user-community building and support. We currently lack the resources to undertake these tasks.

Adequately planned, resourced and implemented software frameworks are potentially important enablers of improved health economic modelling practice. Reference models \citep{Afzali2013}, methodological innovation to improve model transferability \citep{craig2018taking} and the development of regularly updated CHEMs that can support ``living Health Technology Assessment'' \citep{thokala2023living} would all be facilitated by a trusted software framework for implementing more transparent, reusable and updatable CHEMs. A future software framework for CHEMs would ideally incorporate a base set of features useful to developers of computational models across all domains of public health and health economics, with the capability for community-led extensions that are tailored to the needs of modellers focused on specific health-conditions. Such approaches will likely be important to ensure a sufficient potential user-base to make CHEM software frameworks sustainable.

\hypertarget{conclusion}{%
\section{Conclusion}\label{conclusion}}

Our software framework provides a solution to some technical barriers we faced when seeking to implement a transferable CHEM in youth mental health. Our framework can be used as a prototype for developing future software frameworks that address the user-requirements of the broader health economic modelling community.

\hypertarget{acknowledgement}{%
\subsection*{Acknowledgement}\label{acknowledgement}}
\addcontentsline{toc}{subsection}{Acknowledgement}

The authors would like to acknowledge the contribution of John Gillam who provided advisory input to this research.

\hypertarget{availability-of-data-and-materials}{%
\subsection*{Availability of data and materials}\label{availability-of-data-and-materials}}
\addcontentsline{toc}{subsection}{Availability of data and materials}

The ready4 framework is documented at \url{https://www.ready4-dev.com}. Documentation on readyforwhatsnext is available at \url{https://readyforwhatsnext.org/}. Development versions of all code repositories referenced in this article are available in \url{https://github.com/ready4-dev/}. Archived code releases are available in \url{https://zenodo.org/communities/ready4}. All data repositories referenced in this article are available in \url{https://dataverse.harvard.edu/dataverse/ready4}.

\hypertarget{ethics-approval}{%
\subsection*{Ethics approval}\label{ethics-approval}}
\addcontentsline{toc}{subsection}{Ethics approval}

Software framework development did not involve human subject research and was not ethically reviewed. The utility mapping modules were originally developed to implement a previously reported study that was reviewed and granted approval by the University of Melbourne's Human Research Ethics Committee, and the local Human Ethics and Advisory Group (1645367.1).

\hypertarget{funding}{%
\subsection*{Funding}\label{funding}}
\addcontentsline{toc}{subsection}{Funding}

Software framework development was funded by Orygen, VicHealth, Victoria University and an Australian Government Research Training Program (RTP) Scholarship to Matthew Hamilton. The utility mapping study to which utility mapping modules were applied was funded by the National Health and Medical Research Council (NHMRC, APP1076940), Orygen and headspace.

\hypertarget{conflict-of-interest}{%
\subsection*{Conflict of Interest}\label{conflict-of-interest}}
\addcontentsline{toc}{subsection}{Conflict of Interest}

None declared.

\hypertarget{author-contributions}{%
\subsection*{Author contributions}\label{author-contributions}}
\addcontentsline{toc}{subsection}{Author contributions}

MPH led the conceptualisation and software implementation of ready4 and readyforwhatsnext and manuscript drafting. MPH, CG and GW co-authored the software described in this article. All co-authors critically reviewed and revised manuscript drafts and approved the final manuscript text.

\newpage
\appendix
\counterwithin{figure}{section}
\counterwithin{table}{section}

\bibliography{Preprint.bib}

\end{document}